# Root approach for estimation of statistical distributions

Yu. I. Bogdanov[*a,b,c], N.A. Bogdanova[b]

[a]Institute of Physics and Technology, Russian Academy of Science; [b]National Research University of Electronic Technology MIET; [c]National Research Nuclear University 'MEPHI'

## ABSTRACT

Application of root density estimator to problems of statistical data analysis is demonstrated. Four sets of basis functions based on Chebyshev-Hermite, Laguerre, Kravchuk and Charlier polynomials are considered. The sets may be used for numerical analysis in problems of reconstructing statistical distributions by experimental data. Based on the root approach to reconstruction of statistical distributions and quantum states, we study a family of statistical distributions in which the probability density is the product of a Gaussian distribution and an even-degree polynomial. Examples of numerical modeling are given.

The results of present paper are of interest for the development of tomography of quantum states and processes.

**Keywords:** root density estimator, statistical data analysis, quantum informatics

## 1. ROOT APPROACH

Quantum informatics is a modern, rapidly evolving field of science and technology, which is based on application of quantum systems for realization of principally new methods of data transmission and computation (quantum communications, quantum cryptography, quantum computers) [1, 2].

There are a number of serious obstacles to the construction of efficient systems of quantum information processing. One of them is "fragility" and "intangibility" of a quantum state, which is the main object of quantum physics and quantum informatics. A quantum state is described by a so-called state vector, which is a complex vector in abstract Hilbert space that describes probability amplitudes for observing various basis states. The state vector is a fundamentally different information carrier from its classical analogs. One important distinction of quantum systems from classical ones is the fundamental need for statistical description of their behavior. A measurement upon a quantum object changes its quantum state (wave function reduction). This leads to the necessity for statistical (ensemble) approach: while the act of measurement destroys the quantum state of a single object, there is a whole ensemble of the objects.

A quantum registry, which includes $n$ quantum bits (qubits) is described by a state vector which has $2^n$ complex numbers. Statistically, this implies that control of a quantum system can be described by a multiparametric problem of reconstructing a quantum state of statistical ensemble from measurements upon its representatives.

As a practical matter, the multiparametric problem of reconstruction of quantum states plays an important role in realizing the tasks faced by a developer of quantum informational systems. These interconnected tasks are - generation of quantum systems in given quantum states, their transformation during transmission by quantum channels or during quantum computations and observing (measuring) the output of the system. The ability to reconstruct quantum states provides a basis for solving such problems as fine tuning of quantum informational systems, control of precision and functioning stability, detection of outside intrusions to the system.

Multiparametric statistical estimation of quantum states is also clearly interesting in fundamental sense, because it provides a tool for analysis of such basic notions of quantum theory as the statistical nature of its predictions, the superposition principle, Bohr's complementarity principle, etc.

Among all the possible methods of reconstruction of quantum states the most important are those that allow for the precision of estimation close to the theoretically feasible one in problems of high dimension. Construction of such estimators based on traditional methods of mathematical statistics leads to computational difficulties that quickly become insolvable as the dimension of the problem grows. The model which stands out is the so-called root model in which the structure of statistical theory is a-priori consistent with the structure of probabilities in quantum mechanics [3, 4].

---

[*] e-mail: bogdanov_yurii@inbox.ru

An application of the root approach to the problems of quantum tomography and quantum cryptography allowed us to experimentally prove in the joint works between Institute of Physics and Technology and Moscow State University the possibility of reconstruction of quantum states with the precision greatly exceeding that in other works [5-11].

The proposed approach is based on a "symbiosis" of mathematical apparatus of quantum mechanics and the Fisher's maximum likelihood principle in order to get multiparameteric asymptotically effective estimation of density as well as quantum states with the most simple and fundamental properties.

The new method is based on representation of probability density as a square of absolute value of some function (called psi-function by analogy with quantum mechanics). The psi-function is presented as a decomposition by an orthonormal set of functions. The coefficients of decomposition are estimated by the maximum likelihood method.

The introduction of a psi-function as a mathematical tool for statistical data analysis implies that instead of density of distribution one considers the square root of it, i.e.:

$$p(x) = |\psi(x)|^2. \tag{1}$$

It is assumed that the psi-function depends on $s$ unknown parameters $c_0, c_1, ..., c_{s-1}$ that are coefficients of decomposition by some set of orthonormal basis functions:

$$\psi(x) = \sum_{i=0}^{s-1} c_i \varphi_i(x). \tag{2}$$

The maximum likelihood principle implies that the "most likely" estimators of unknown parameters $c_0, c_1, ..., c_{s-1}$ are the values that maximize the likelihood function and its logarithm.

$$\ln L = \sum_{k=1}^{n} \ln p(x_k | c) \to \max. \tag{3}$$

Here $x_1, ..., x_n$ – is a sample of size $n$.

The condition of extreme value of the logarithmic likelihood function and the normalization condition lead to the following likelihood:

$$\frac{1}{n} \sum_{k=1}^{n} \left( \frac{\varphi_i^*(x_k)}{\sum_{j=0}^{s-1} c_j^* \varphi_j^*(x_k)} \right) = c_i, \quad i = 0, 1, ..., s-1. \tag{4}$$

The likelihood equation has a simple quasi-linear structure and allows for construction of an effective, quickly converging iterational procedure (for instance when the number of estimated parameters lies in tens, hundreds or even thousands) [3, 4]. This is how the considered problem differs from other known problems solved by the maximum likelihood method, such as with the growth in the number of estimated parameters the complexity of numerical analysis quickly grows, while the stability of the algorithms steeply falls.

Above we have considered the case of a continuous distribution. The case of a discrete distribution is described similarly.

Let us present some basic sets of basis functions, which are convenient to use for numerical analysis in the problems of reconstruction of statistical distributions by experimental data.

The basis orthonormal set of Chebyshev-Hermite functions has the following form:

$$\varphi_k(x) = \frac{1}{\left(2^k k! \sqrt{\pi}\right)^{1/2}} H_k(x) \exp\left(-\frac{x^2}{2}\right), \quad k = 0, 1, 2, \ldots. \tag{5}$$

Here $H_k(x)$ – is the Chebyshev-Hermite polynomial of *k*-th order.

The basis describes stationary states of a quantum harmonic oscillator.

The Chebyshev-Hermite basis is particularly convenient for the following reason. If in the zero-order approximation the distribution is considered to be Gaussian, then one simply assigns the value of the ground state to the state vector, while deviations from Gaussian distribution are modeled by adding higher order harmonics. Note that the final distribution may greatly differ from the zero-order approximation (it can be asymmetric, multi-modal etc.)

The considered set of basis functions can be effectively and precisely applied to the problems of reconstruction of arbitrary statistical distributions defined on the whole line ($-\infty < x < +\infty$).

For distributions defined on a half-line (say $0 \leq x < +\infty$), it is more convenient to use another basis set of orthonormal functions which is based on Laguerre polynomials. The corresponding basis functions have the following form:

$$\varphi_k(x) = L_k(x) \exp\left(-\frac{x}{2}\right), \quad k = 0, 1, 2, \ldots. \tag{6}$$

Here $L_k(x)$ – is Laguerre polynomial of *k*-th order.

In the considered set of basis functions the ground state corresponds to exponential distribution. Accounting for higher harmonics in the psi-function decomposition describes deviations of the distribution from the exponential one.

The two principal models of discrete distributions of mathematical statistics are binomial and Poisson distributions. In the root approach these distributions act as zero-order approximations for reconstruction of discrete distributions of general form.

An orthonormal set of basis functions based on Kravchuk polynomials allows one to describe binomial type multiparametric distributions.

The corresponding discrete distribution is defined at points $x = 0, 1, 2, \ldots, N$. By analogy to the ordinary binomial distribution, we may call the random number $x$ the number of "successes" in a series of $N$ independent "experiments". At zero-order approximation the distribution under consideration is an ordinary binomial distribution. The considered basis functions are defined as:

$$\varphi_k(x) = \left(\frac{k!(N-k)!}{(pq)^k} \frac{p^x q^{N-x}}{x!(N-x)!}\right)^{1/2} K_k^p(x), \quad k = 0, 1, 2, \ldots, N. \tag{7}$$

Here $p$ – is the parameter corresponding to the average probability of "success", $q = 1 - p$, $K_k^p(x)$ – Kravchuk polynomial of *k*-th order corresponding to a given $p$.

A basis orthonormal set of functions based on Charlier polynomials allows one to model Poisson-type multiparametric distributions. Corresponding distributions are defined on non-negative whole number points $x = 0, 1, 2, \ldots$. At zero-order approximation the considered distribution is an ordinary Poisson distribution. Its basis functions have the form:

$$\varphi_k(x) = \left(\frac{\lambda^{k+x} e^{-\lambda}}{k! x!}\right)^{1/2} C_k^\lambda(x), \quad k = 0, 1, 2, \ldots. \tag{8}$$

Here $\lambda$ – is the parameter corresponding to the average number of "successes" (average value of $x$), $C_k^\lambda(x)$ – is the Charlier polynomial of $k$-th order corresponding to a given $\lambda$.

## 2. ROOT APPROACH

To illustrate the method presented above, we present examples of reconstructions of statistical distributions. Consider first the examples corresponding to a real-valued wave function

Let us look at Figure 1 and Figure 2. One can note very close correspondence (almost exact) between theoretical distributions and root estimators. For comparison, traditional methods' results are also presented, which clearly lack precision. The presented research demonstrates significant superiority of the root approach over kernel estimators of Rosenblatt-Parsen and projection estimators of Chentsov for the problems of reconstruction of statistical distributions.

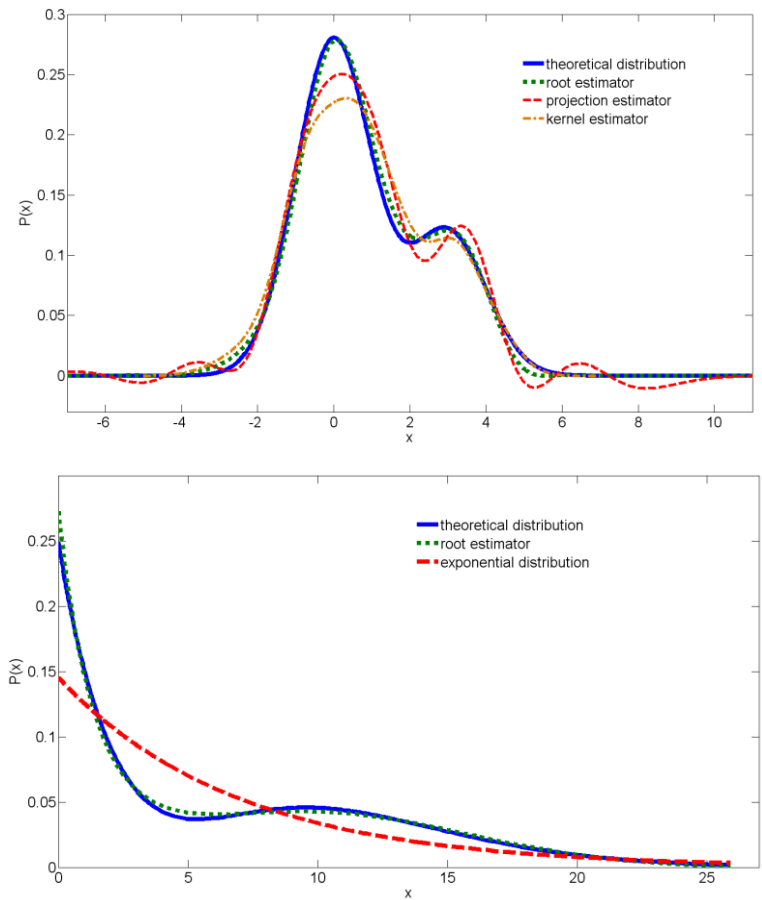

Figure1. Examples of reconstruction of continuous statistical distributions. The upper picture: mixture of two normal distributions with parameters: $\sigma_1 = \sigma_2 = 1$, $\mu_1 = 0$ (70%), $\mu_1 = 3$ (30%). Sample size 200. Comparison of root estimation using Chebyshev-Hermite polynomials basis with kernel and projection estimators. The lower picture: equally- weighted mixture of exponential distribution with mean equal to 2 and chi-squared distribution with 12 degrees of freedom. Sample size 400. Solid line – theoretical density, dot line – root estimation using Laguerre polynomials basis, dashed line – exponential estimation.

For the distribution represented in Figure 1 above, 100 independent numerical experiments were performed. The results are shown in Table 1.

We characterize the quality of the approximation in the $L_1$-norm by the difference between the exact density and its estimation.

$$\Delta = \int |\hat{p}(x) - p(x)| dx. \qquad (9)$$

Table 1. Comparison of the accuracy of the estimators (100 numerical experiments)

| Estimator | Root | Projection | Kernel |
|---|---|---|---|
| Mean value of $\Delta$ | 0.0967 | 0.1372 | 0.1296 |
| Standard deviation of $\Delta$ | 0.0310 | 0.0481 | 0.0330 |

Results presented in Table 1 show that the root density estimator has a significant advantage over other methods.

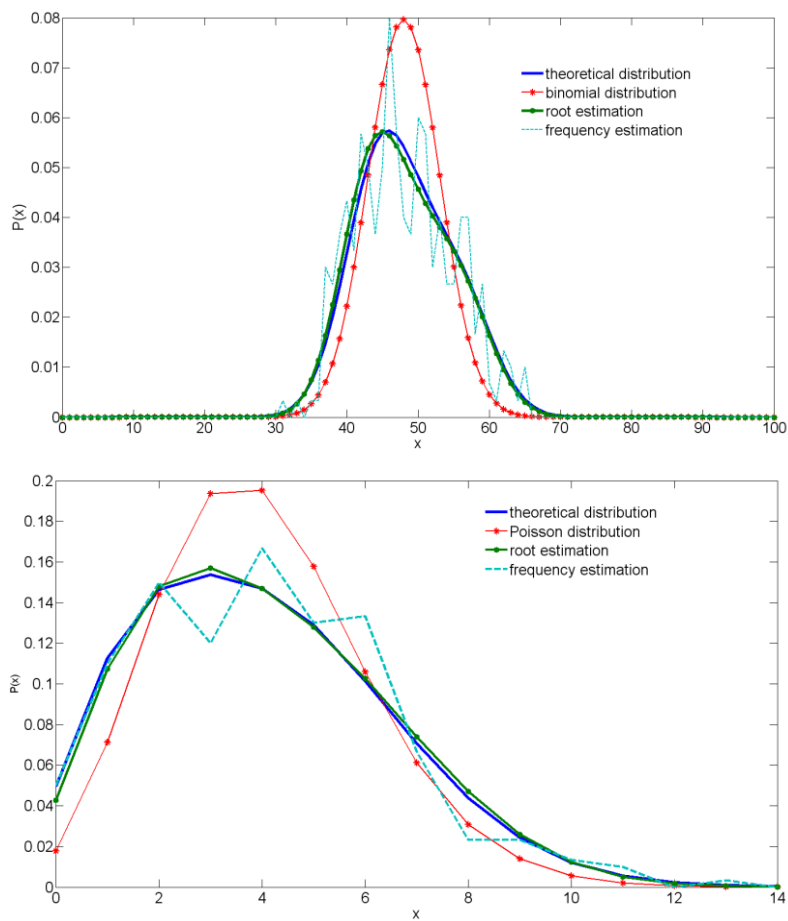

Figure2. Examples of reconstruction of discrete statistical distributions. The upper picture: 2-to-1 weighted mixture of binomial distributions with parameters $N_1 = N_2 = 100$, $p_1 = 0.45$ and $p_2 = 0.55$ correspondingly. Sample size 300. Solid line – theoretical probability distribution, dot line – root estimation with Kravchuk polynomials basis, stars – binomial estimation, dashed line – frequency estimation. The lower picture: 1-to-2 weighted mixture of Poisson distributions with parameters $\lambda_1 = 2$ and $\lambda_2 = 5$ correspondingly. Sample size 300. Solid line – theoretical probability distribution, dot line – root estimation with Charlier polynomials basis, stars – Poisson estimation, dashed line – frequency estimation.

We now consider examples corresponding to a complex-valued wave function.

Consider the family of statistical distributions in which the density is the product of a Gaussian distribution and a polynomial:

$$P(x) = f(x)\exp(-x^2). \tag{10}$$

Here, $f(x)$ is a polynomial. Gaussian distribution in (10) corresponds to the variance equal to 0.5, and stands for the ground state of the harmonic oscillator.

The coefficients of the polynomial $f(x)$ must be selected in such ways, so that the density is non-negative. Furthermore, the distribution must satisfy the normalization condition. It is not difficult to prove that any such density can be represented as the square modulus of a wave function, which is complex-valued in general:

$$P(x) = |\psi|^2. \tag{11}$$

Indeed, the density can be nonnegative everywhere only if the polynomial $f(x)$ has an even degree. Let this degree be equal to $2n$. According to the fundamental theorem of algebra, the polynomial has exactly $2n$ roots. The coefficients of the polynomial $f(x)$ are real-valued, therefore all complex roots must occur in pairs (if $z_j$ is a root, then the complex conjugate $z_j^*$ is also a root). Some of the roots may be real-valued, but they must be an even multiple (2, 4, 6, etc.). Such real-valued roots can also be clearly divided into pairs $z_j$ and $z_j^*$, where simply $z_j^* = z_j$ for real numbers. If any real-valued root had odd multiplicity (1, 3, 5, etc.), then the polynomial would not have a definite sign near the root, which would lead to an unacceptable negative density.

Let us consider the polynomial in the form:

$$f(x) = a_0 + a_1 x + a_2 x^2 + \ldots + a_{2n} x^{2n}. \tag{12}$$

Clearly, because of the non-negativity of the density, the following relation always holds $a_{2n} > 0$. Having calculated the roots of the polynomial, we can write:

$$f(x) = a_{2n} \prod_{j=1}^{n} (x - z_j)(x - z_j^*). \tag{13}$$

Then, it is straightforward to construct all possible psi-functions

$$\psi(x) = \sqrt{a_{2n}} \exp\left(-\frac{x^2}{2}\right) \prod_{j=1}^{n} (x - y_j). \tag{14}$$

Here as the value $y_j$ we can select, at our discretion, the root $z_j$ or the complex conjugate root $z_j^*$. If for the function $\psi(x)$ $z_j$ has been selected, then for $\psi^*(x)$ $z_j^*$ will automatically be selected, and vice versa.

Psi-function (14) is obviously the product of a Gaussian function and a complex polynomial of degree $n$. Let $s$ be the dimension of the Hilbert space (number of functions in Chebyshev-Hermite basis). The corresponding functions are $\varphi_j(x)$, $j = 0, 1, 2, \ldots, s-1$. They contain polynomials from zero degree to degree $s-1$ inclusively. If $s - 1 = n$, then a basis set will be suitable for constructing the wave functions discussed above.

Figure 3 corresponds to the case of a three-dimensional Hilbert space ($s = 3$), where we have 4 unknown parameters for the density distribution. The sample size is 1000. The density distribution is the product of a Gaussian distribution and a polynomial of the fourth degree. The presented graph illustrates the advantage of the root density estimator. The reconstruction errors $\Delta$ in $L_1$ norm for the root, kernel and projection estimators respectively, have the following values: 0.03363; 0.08113; 0.1031.

We have performed 100 numerical experiments, which are similar to the case shown in Figure 3. The results are presented in Table 2.

Table 2. Comparison of the accuracy of the estimators (100 numerical experiments), s=3

| **Estimator** | **Root** | **Projection** | **Kernel** |
|---|---|---|---|
| Mean value of $\Delta$ | 0.04802 | 0.09489 | 0.09729 |
| Standard deviation of $\Delta$ | 0.02100 | 0.01981 | 0.01675 |

Results presented in Table 2 show that the root density estimator has a radical advantage over other methods. In particular, the root estimator was the best one in all of the 100 numerical experiments.

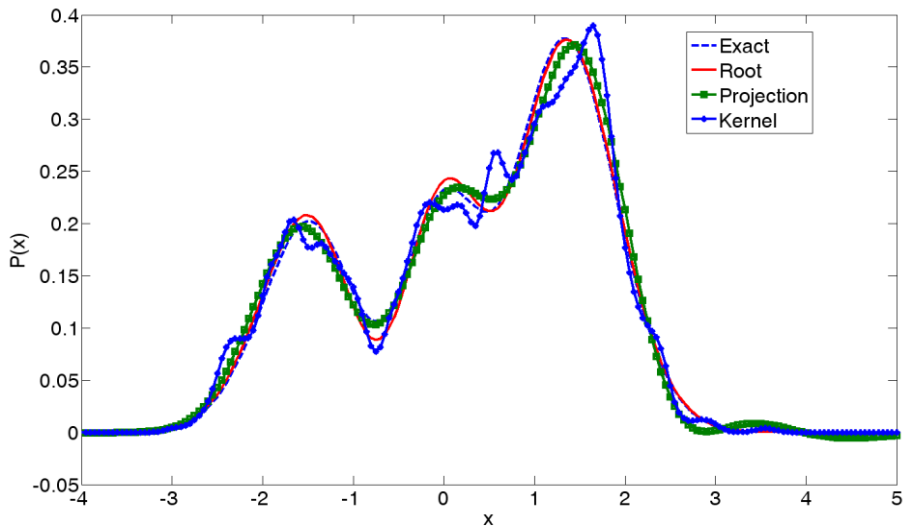

Figure 3. Comparison of root estimator based on complex-valued wave function, with the projection and kernel estimators (three-dimensional Hilbert space – $s = 3$).

Figure 4 is similar to Figure 3, but now $s = 4$ (four-dimensional Hilbert space)

Reconstruction errors $\Delta$ in $L_1$ norm for the root, kernel and projection estimators respectively, have the following values: 0.02967; 0.1416; 0.09293.

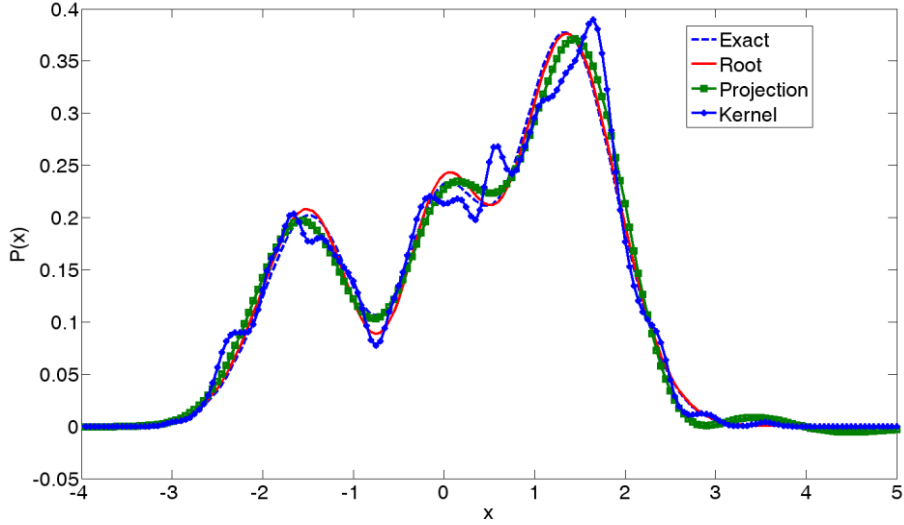

Figure 4. Comparison of root estimator based on complex-valued wave function, with the projection and kernel estimators (four-dimensional Hilbert space – $s=4$).

We have performed 100 numerical experiments, which are similar to the case shown in Figure4. The results are presented in Table 3.

Table 3. Comparison of the accuracy of the estimators (100 numerical experiments), s=3

| Estimator | Root | Projection | Kernel |
| --- | --- | --- | --- |
| Mean value of $\Delta$ | 0.05347 | 0.1590 | 0.1079 |
| Standard deviation of $\Delta$ | 0.01656 | 0.01498 | 0.01663 |

As previously, the root estimator was the best one in all of the 100 numerical experiments.

## 3. CONCLUSIONS

We examined the root approach to the reconstruction of statistical distributions. A significant advantage of the root approach in comparison with the methods of Rosenblatt-Parzen and Chentsov has been demonstrated.

We examined the family of statistical distributions in which the density was the product of a Gaussian distribution and a polynomial. The peculiarity of this problem is that it is necessary to use the complex-valued wave functions for solution of classical (non-quantum) tasks.

The results of present paper are of interest for the development of tomography of quantum states and processes. From a mathematical point of view, however, it is important also that the root approach defines a general approach to solve not only the problems of quantum tomography, but also problems of statistical data analysis.

## ACKNOWLEDGEMENTS

This work was supported in part by Russian Foundation of Basic Research (project 13-07-00711), and by the Program of the Russian Academy of Sciences in fundamental research.